# Unconventional superconductivity of NdFeAsO$_{0.82}$F$_{0.18}$ indicated by the low temperature dependence of the lower critical field $H_{c1}$

X.L. Wang[1*], S. X. Dou[1]

[1] *Institute for Superconducting and Electronic Materials, University of Wollongong, Wollongong, New South Wales 2522, Australia*

Zhi-An Ren[2], Wei Yi[2], Zheng-Cai Li[2], Zhong-Xian Zhao[2]

*National Laboratory for Superconductivity, Institute of Physics, Chinese Academy of Sciences, P. O. Box 603, Beijing, P. R. China, 100190*

Sung-IK Lee[3**]

[3]*National Creative Research Initiative Center for Superconductivity, Department of Physics, Sogang University, Seoul 121-742, Republic of Korea*

We measured the initial M-H curves for a sample of the newly discovered superconductor NdFeAsO$_{0.82}$Fe$_{0.18}$, which had a critical temperature, $T_c$, of 51 K, and was fabricated at the high pressure of 6 GPa. The lower critical field, $H_{c1}$, was extracted from the deviation point of the Meissner linearity in the M-H curves, which show linear temperature dependence in the low temperature region down to 5 K. The $H_{c1}$(T) indicates no s-wave superconductivity, but rather an unconventional superconductivity with a nodal gap structure. Furthermore, the linearity of $H_{c1}$ at low temperature does not hold at high temperature, but shows other characteristics, indicating that this superconductor might have multi-gap features. Based on the low temperature nodal gap structure, we estimate that the maximum gap magnitude $\Delta_0 = (1.6 \pm 0.2)$ k$_B$T$_c$.

*Email: xiaolin@uow.edu.au

**Email: silee77@sogang.ac.kr

The recent discovery of superconductivity in rare earth iron-based compounds has lead to intensive theoretical and experimental activity in the superconductivity community [1-6]. One of the major issues with these compounds is why superconductivity appears at such high temperatures. Another issue is what type of gap symmetry could directly lead to the superconducting pairing mechanism [7]. Up to now, the nature of the gap symmetry is still unclear, because there are still conflicting results on this quantity, such as whether there is one gap with or without a node. Further complications come from the claims for multi-gap features in the new superconductors. For example, recent tunneling measurements show a Bardeen Cooper Schrieffer (BCS) s-wave like gap in the superconductor $SmFeAsO_{0.85}F_{0.15}$ [7], while the electronic specific heat coefficient in the low temperature limit shows a nodal gap structure [8]. Nuclear magnetic resonance (NMR) measurements in $PrFeO_{0.89}F_{0.11}$ (critical temperature $T_c$ = 45 K) shows that Cooper pairs are in the spin-singlet state, with two energy gaps opening below $T_c$ [9].

It is therefore very important to further investigate gap symmetry in the new class of superconductors. One of the simple methods to determine the gap symmetry is the measurement of the penetration depth from the lower critical field, $H_{c1}$. The low temperature $H_{c1}$ can be obtained from the simple initial M-H curves. Compared to the other physical quantities, penetration depth is a rather good quantity for studying the gap symmetry, because it is rather bulk sensitive and insensitive to the small details of the surface conditions. Up to now, there have been almost no reports of $H_{c1}$ measurements on REFeAsO based superconductors, referred to as RE-1111, where RE is a rare earth element, except for F-doped La-1111, $Sm_{0.95}La_{0.05}FeAsO_{0.85}F_{0.15}$ ((Sm,La)-1111), and one on K doped $Ba_{0.6}K_{0.4}Fe_2As_2$ (referred to as Ba-122) [10-12] single crystals. However, the gap features from the $H_{c1}$ measurements for both Ba-122 and (Sm,La)-1111 phases show rather different results. For La-1111, $H_{c1}$ shows a clear linear T dependence down to 2 K, which indicates unconventional superconductivity and a nodal gap function with maximum gap magnitude $\Delta_0$ = 4.0 $\pm$ 0.6 meV. However, for the Ba-122, the gap symmetry from $H_{c1}$ is quite different from that of La-1111 and (Sm,La)-1111. For the Ba-122 case, the data on $H_{c1}(T)$ show two gap features, with a small gap of $\Delta_a(0) = 2.0 \pm 0.3$ meV, and a large gap of $\Delta_b(0) = 8.9 \pm 0.4$ meV. Also, for the Ba-122 case, the in-plane superfluid density is quite large, which indicates breakdown of the Uemura plot, while RE-1111 fits the Uemura plot. Although many Fe-As based superconductors have been fabricated, research progress on the gap symmetry from the $H_{c1}$ measurements has been slow, while the issue of the gap symmetry is still unsettled by the $H_{c1}$ measurements. To resolve the conflicting evidence on the gap symmetry from the $H_{c1}$ measurements, it is urgently needed to accumulate more data on the Fe-As superconductors. Up to now, there has been no report of $H_{c1}$ measurements on Nd-1111 superconductors, which have a $T_c$ higher than 50 K.

In this letter, we present the results of our $H_{c1}$ measurements of an Nd-1111 sample with $T_c$ = 51 K. We obtained the penetration depth from the $H_{c1}$ measurements for the full range of temperatures below the transition temperature.

The $H_{c1}(T)$ indicates no s-wave superconductivity, but rather unconventional superconductivity with a nodal gap structure. Furthermore, the linearity of $H_{c1}$ at low temperature does not hold at high temperature, where it shows different features, indicating that this superconductor might have multi-gap characteristics. Our observation is in strong contrast to what has been reported for La-1111, Sm,La-1111, and Ba-122, which show either one nodal gap or an s-wave multi-gap feature.

The sample used in the present study has the nominal composition of $NdFeAsO_{0.82}Fe_{0.18}$ with high phase purity and $T_c$ of 51 K. The sample was synthesized at 1200 °C under 6 GPa. The details of sample fabrication conditions and phase analysis can be found in Refs. [3,13]. Magnetization measurements were performed by using a magnetic properties measurement system (MPMS; Quantum Design). A series of virgin M-H curves were measured between 5 and 48 K with an interval of 2 K.

The virgin M-H curves measured at different temperatures are shown in Fig. 1. The magnetic field $H_{c1}^*$ at which the magnetization starts to deviate from linearity, i.e. the field starts to penetrate into the sample, was carefully analyzed. However, the $H_{c1}^*$ is not the same as the real lower critical field, $H_{c1}$, due to the geometric effect. We used a well adapted relation between $H_{c1}^*$ and $H_{c1}$, proposed by Brandt [14]: $H_{c1} = H_{c1}^*/\tanh(\sqrt{0.36b/a})$, where a and b are the width and thickness of a plate-like superconductor, respectively. We used $b/a = 10$ for our case. As claimed by Ren et al., [10], we can neglect the surface barrier effect in our $NdFeAsO_{0.82}F_{0.18}$. In this work, in order to analyze the $H_{c1}$ of our Nd-1111 superconductor, we adopted the same analysis method that was used by Ren et al. [10] for analyzing their La-1111 phase compound with $T_c$ of 26 K.

The obtained $H_{c1}$ results against temperature are shown in Fig. 2. It can be seen that the features of $H_{c1}(T)$ for T < 20 K are different from the case where T > 20 K. The $H_{c1}$ is linear with temperature for T < 20 and deviates strongly from linearity above 20 K. This observation on the $H_{c1}(T)$ in our $NdFeAsO_{0.82}F_{0.18}$ sample is quite different from what have been reported for La-1111 and Ba-122 single crystals. In La-1111, the $H_{c1}(T)$ is generally linear over a wide range of temperatures from 2 K up to $T_c$. [10] On the other hand, for Ba-122 single crystal, a pronounced kink was observed at T = 15 K, below which the $H_{c1}$ tends to saturate at very low temperature [12].

We calculated the penetration depth $\lambda$ from the relation of $H_{c1} = \phi_0/4\lambda^2 (\ln \kappa +0.5)$, where $\phi_0 = hc/2e = 2.07 \times 10^{-7}$ Oe cm$^2$ is the flux quantum and $\kappa$ is the Ginzburg-Landau (G-L) parameter. The temperature dependence of $\lambda$ is depicted in Fig. 3. Assuming $\kappa$ is 100, the $\lambda(0)$ calculated from this formula is about 320 nm, which is very similar to that obtained for (Sm,La)-1111 [11]. $\lambda$ for our Nd-1111 increases linearly from 320 nm for T = 0 K to 600 for T = 25 K, and then rapidly rises up to 1400 for T = 40 K. It is clear that the penetration depth of our Nd-1111 shows quite different behavior compared to the penetration depths of La-1111 and Ba-122 single crystal. For our case, we have two clear features: one is the linear temperature dependence of the penetration depth at low temperature, and the other is the behavior at high temperature, which has not been observed in other 1111 type Fe-As superconductors.

It should be pointed out that $\lambda$ can be calculated based on the relations: $\Delta\lambda = \lambda(T) - \lambda(0) = \lambda(0) \sqrt{\pi\Delta_0/2T} \exp(-\Delta_0/2T)$ for isotropic s-wave superconductors, in which the increase in $\lambda$ with temperature is very slow, but this does not agree with our measurement data, as indicated by the closed circles in Fig. 3. For example, if we take the BCS value of $2\Delta_0/k_bT_c = 3.5$, then we would have a BCS s-wave graph in Figure 3. Obviously, s-wave gap symmetry fails to explain our experimental observations.

It is well known that if $\Delta\lambda$ at low temperature is linearly dependent on temperature, then the penetration depth $\lambda(T)$ can be explained by d-wave superconductivity. As can be seen in Fig. 3, $\Delta\lambda$ at low temperature is linear with respect to T, which is consistent with what has been seen in La-1111 [10] with $T_c$ of 26 K.

If we assume the formula for d-wave gap symmetry, for $T \ll T_c$, then $1 - \lambda^2(0)/\lambda^2(T) = 2\ln2\, k_BT/\Delta_0$. From this, we could obtain $\Delta_0 = 1.6\, k_BT_c$ [10]. $\Delta_0$ is the maximum d-wave gap. The d-wave character of $\lambda(T)$ is clearly observed for T < 25 K, as shown in Fig. 3. It should be noted that $\lambda(T)$ increases rapidly, indicating that a second gap feature appears.

In summary, the lower critical field $H_{c1}$ was determined for $NdFeAsO_{0.82}Fe_{0.18}$. $H_{c1}$ shows a linear T dependence in the low temperature region, indicating no s-wave superconductivity, but rather unconventional superconductivity with a nodal gap structure. However, this linearity of $H_{c1}$ at low temperature does not hold at high temperature, but shows other characteristics, indicating that this superconductor might have superconducting multi-gap features. Based on the low temperature nodal gap structure, we estimate that maximum gap magnitude of $\Delta_0 = (1.6 \pm 0.2)\, k_B T_c$.

**Acknowledgments**

This work is supported by the Australian Research Council. This work is also supported by Center for Superconductivity from the program of Acceleration Research of the Ministry of Science and Technology/ Korea Science and Engineering Foundation (MOST/KOSEF) and the Special Fund of Sogang University

Figure captions:

Fig. 1 Superconducting initial magnetization measured at various temperatures from 5 up to 48 K, with an interval of 2 K.

Fig. 2 Temperature dependence of $H_{c1}$.

Fig. 3 Temperature dependence of the penetration depth (open circles). The closed circles represent the BCS s-wave superconductor model with $2\Delta_0/k_B T_c = 3.5$.

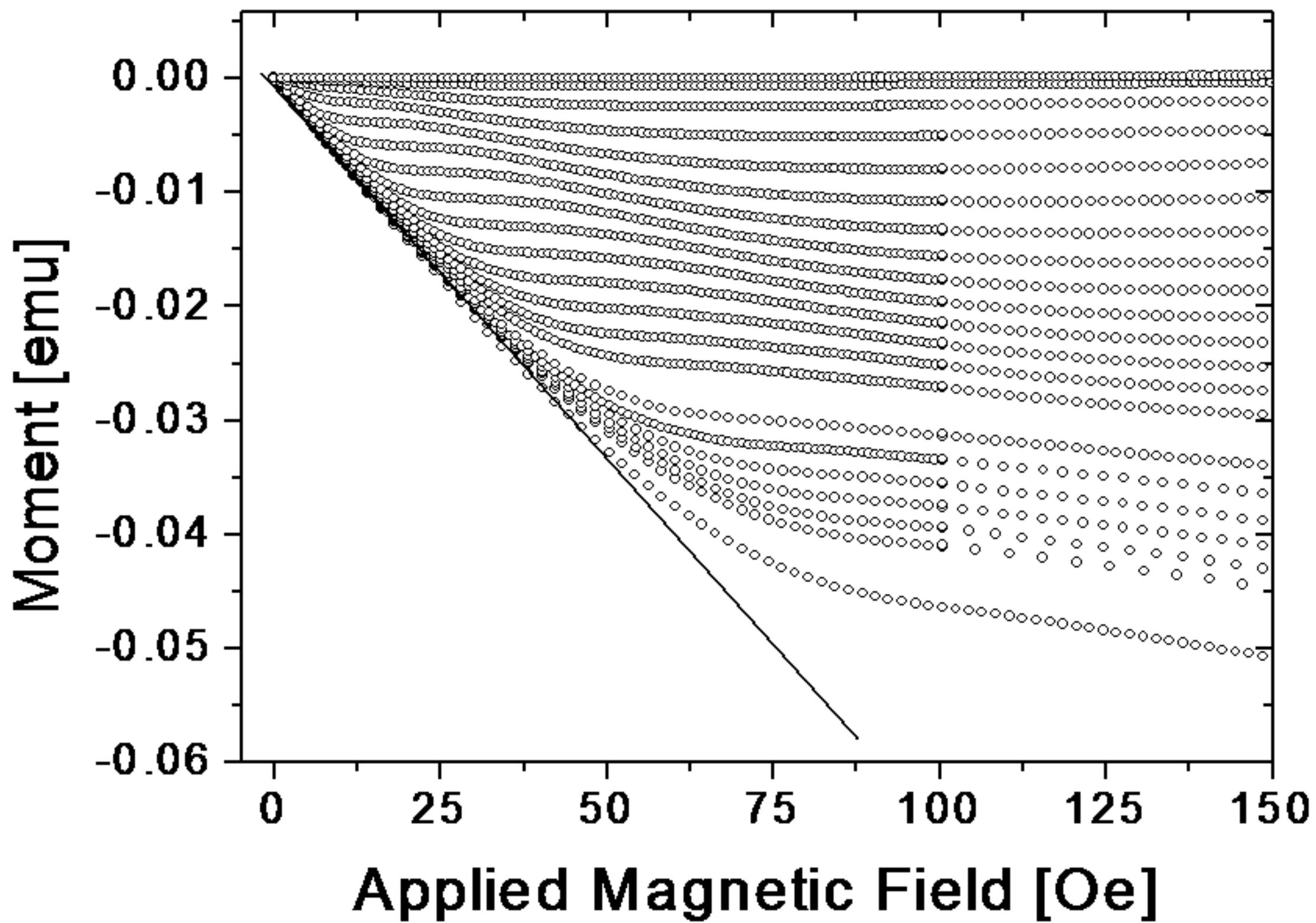

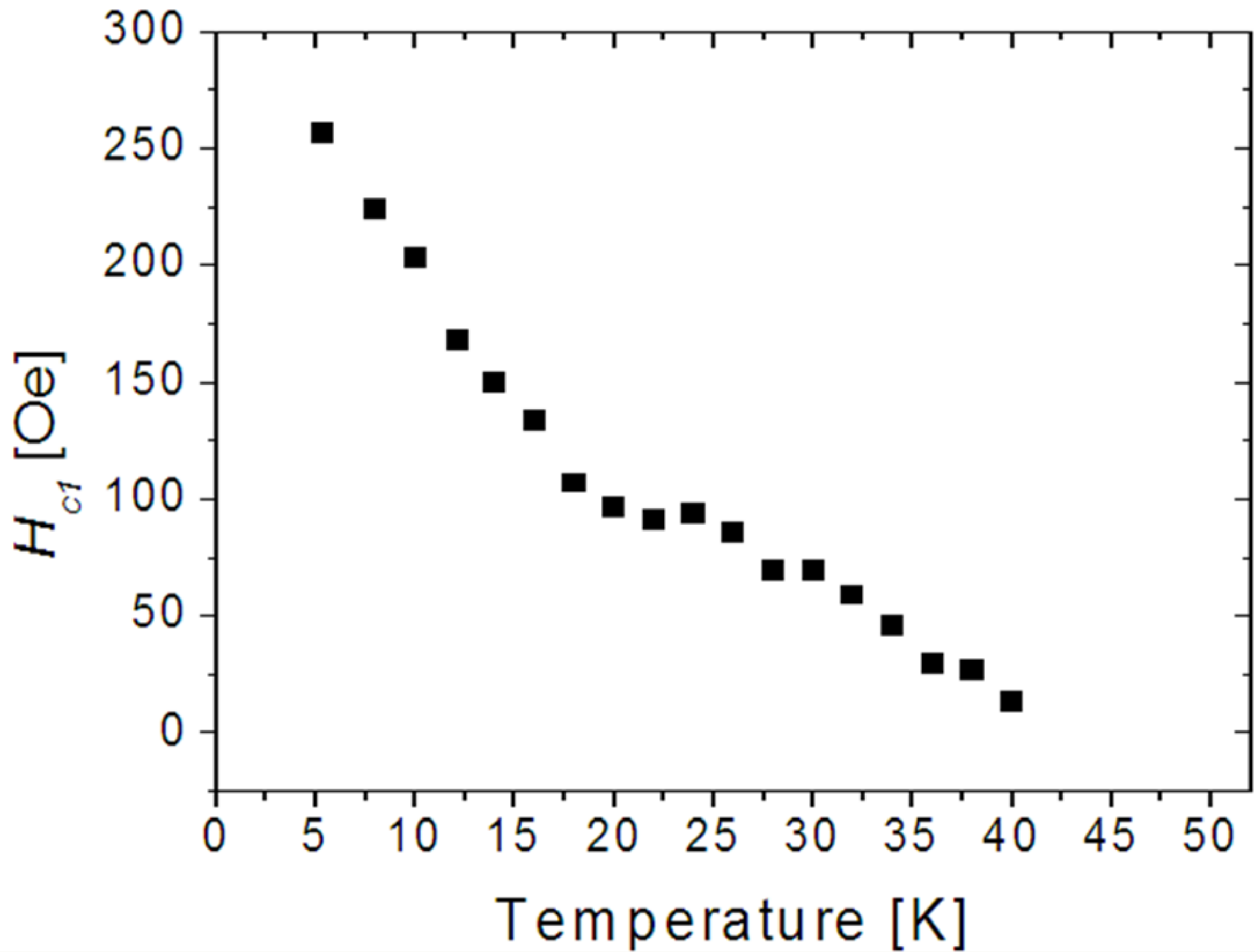

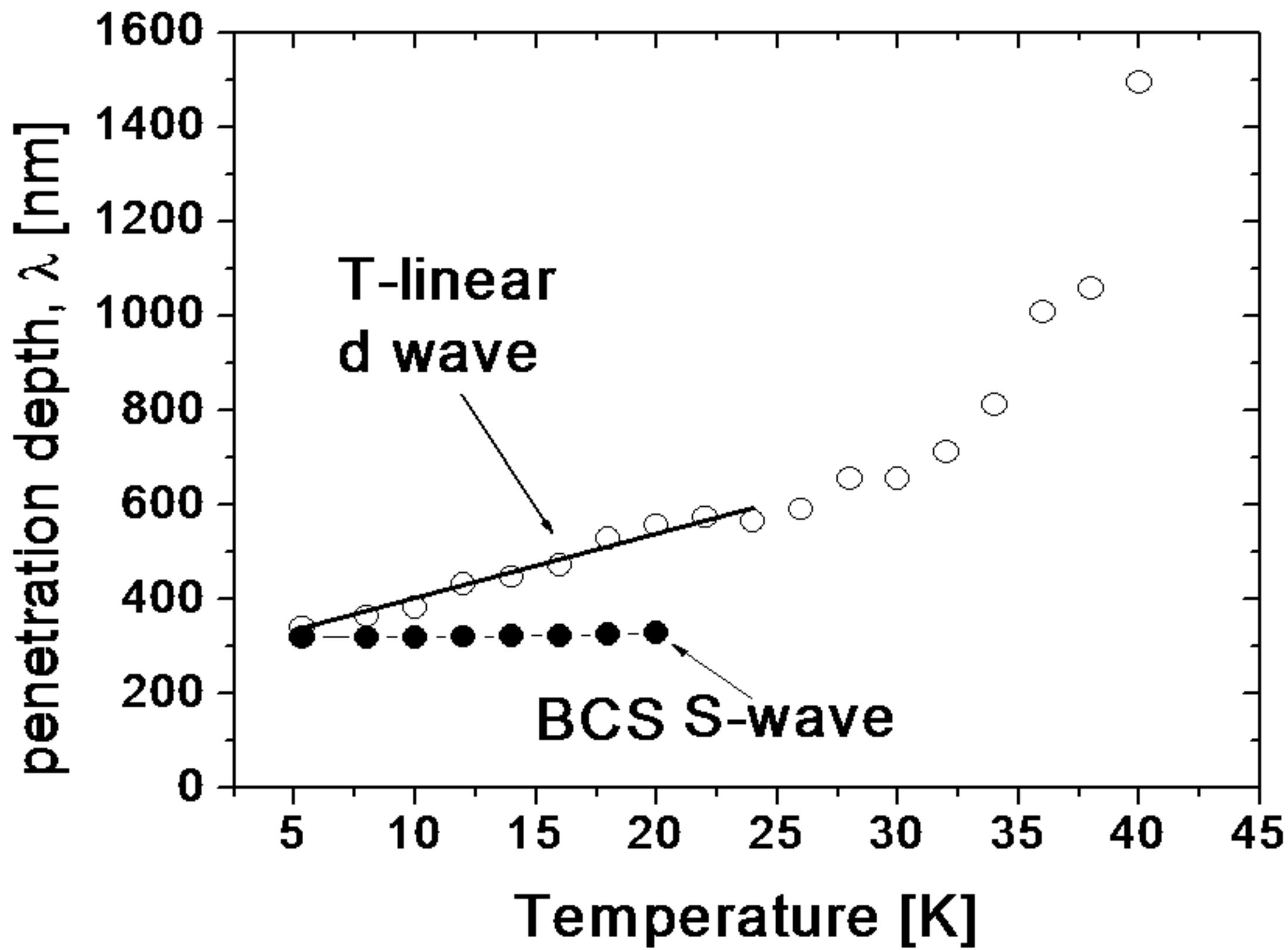